\documentclass{elsart}

\usepackage{amsmath}
\usepackage{graphicx}

\graphicspath{{img/}}

\makeatletter
\def\@date{22 May 2002}
\makeatother

\renewcommand{\l}{\ell}
\renewcommand{\L}{L}%\mathcal{L}}

\begin{document}

\begin{frontmatter}
\title{Laser controlled molecular switches and transistors}
\author{J\"org Lehmann},
\author{S\'ebastien Camalet},
\author{Sigmund Kohler}, and
\author{Peter H\"anggi\corauthref{cor1}}
\corauth[cor1]{Corresponding author. Tel.: +49-821-598-3250; 
fax: +49-821-598-3222; e-mail: Peter.Hanggi@physik.uni-augsburg.de}
\address{Institut f\"ur Physik, Universit\"at Augsburg, 
         Universit\"atsstra{\ss}e~1, D-86135 Augsburg, Germany}

\begin{abstract}
We investigate the possibility of optical current control through single molecules
which are weakly coupled to leads.  A master equation approach for the transport
through a molecule is combined with a Floquet theory for the time-dependent
molecule.  This yields an efficient numerical approach to the evaluation of
the current through time-dependent nano-structures in the presence of a finite
external voltage.  We propose tunable optical current switching in two- and
three-terminal molecular electronic devices driven by properly adjusted laser
fields, \textit{i.e.} a novel class of molecular transistors.

\end{abstract}

\end{frontmatter}

\section{Introduction}

Spurred by the ongoing experimental progress in the field of molecular
electronics \cite{Reed1997a,Cui2001a,Joachim2000a,Schon2001a,Reichert2002a},
the theoretical interest in transport properties of molecules has revived
\cite{Nitzan2001a}.  Tight-binding models for the wire have been used to
compute current-voltage characteristics, within a scattering approach
\cite{Mujica1994a} and from electron transfer theory \cite{Nitzan2001a}.  Both
approaches bear the same essentials \cite{Nitzan2001b}.  For high temperatures,
the wire electrons loose their quantum coherence and the transport is dominated
by incoherent hopping between neighbouring sites \cite{Petrov2001a}.  Recently,
the current-voltage characteristics has been obtained from a quantum-chemical
\textit{ab initio} description of the molecule \cite{Heurich2001a}.  The
results were in good agreement with recent experiments \cite{Reichert2002a}.

Typical electronic excitation energies in molecules are in the range up to an
eV and, thus, correspond to light quanta from the optical and the infrared
spectral regime where most of today's lasers work.  It is therefore natural to
use such coherent light sources to excite molecules and to study their
influence on the transport properties aiming to find a way of manipulating
currents.  One particularly prominent example of quantum control is the
so-called coherent destruction of tunnelling (CDT), \textit{i.e.} the
suppression of the tunnelling dynamics in an \textit{isolated} bistable
potential by the purely coherent influence of an oscillating bias
\cite{Grossmann1991a,Grossmann1992a,Morillo1993a,Goychuk1996a,Grifoni1998a}.
The crucial point there is that the long-time dynamics in a periodically driven
quantum system is no longer dominated by the energies, but rather by the
so-called quasienergies \cite{Grifoni1998a,Shirley1965a,Sambe1973a}.  The
latter may be degenerate for properly chosen driving parameters yielding a
divergent time-scale.  Inspired by these results, we address in this Letter the
question of controlling by use of properly tailored laser fields the transport
through time-dependent \textit{open} systems, \textit{i.e.} systems that allow
for a particle exchange with external leads.

For the computation of electrical currents through wires exposed to strong
laser fields, we put forward Floquet approach \cite{Lehmann2002b}.  The central
idea of this method lies in a non-perturbative solution of the Schr\"odinger
equation of the isolated time-dependent wire plus laser field, while the
wire-lead coupling is treated perturbatively.  The resulting density operator
equation is decomposed into a time-dependent Floquet basis permitting a
numerically efficient treatment.  We generalise here this method to the
analysis of networks with an arbitrary number of contacts to leads.
Subsequently we apply the so obtained formalism to the investigation of optical
current switching in two- and three-terminal devices as prototypical examples
of a new class of molecular transistors.

\section{Model for wire and leads}

%%%% Figure 1 %%%%%%%%%%%%%%%%%%%%%%%%%%%%%%%%%%%%%%%%%%%%%%%%%%%%%%%%%%%%%%%
\begin{figure}
  \begin{center}
    \includegraphics[width=6truecm]{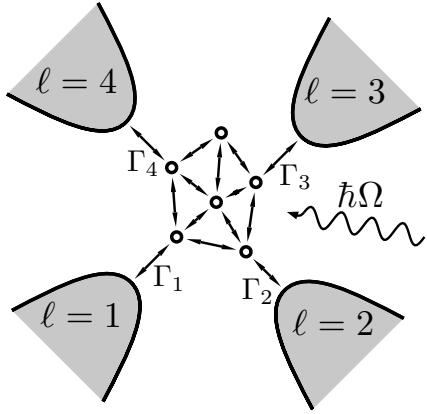}
  \end{center}
  \caption{Molecular circuit consisting of $N=6$ sites of which the
    sites $1,\ldots,L$ are coupled to $\L=4$ leads.}
  \label{fig:system}
\end{figure}
%%%% Figure 1 %%%%%%%%%%%%%%%%%%%%%%%%%%%%%%%%%%%%%%%%%%%%%%%%%%%%%%%%%%%%%%%

We embark by specifying the model Hamiltonian of the entire system as
sketched in Fig.~\ref{fig:system}.  It consists of the molecule in the
laser field, ideal leads, and the molecule-leads coupling Hamiltonian,
\begin{equation}
H(t)=H_{\text{molecule}}(t) + H_{\text{leads}} +
H_{\text{molecule-leads}}\ .
\end{equation}
The irradiated molecule is modelled by a tight-binding description taking into
account $N$ molecular orbitals $|n\rangle$, which are relevant for the
transport.  Disregarding the electron-electron interaction, the most general
form of the Hamiltonian reads
\begin{equation}
\label{eq:Hmolecule} H_{\text{molecule}}(t)=\sum_{n,n'} H_{nn'}(t)\,
c_n^\dagger c_{n'},
\end{equation}
where the fermionic operators $c_n$ and $c_n^\dagger$ destroy and create,
respectively, an electron in the molecular orbital $|n\rangle$. The sums
extend over all tight-binding orbitals.  The $\mathcal{T}$-periodic
time-dependence of the single-particle Hamiltonian
$H_{nn'}(t)=H_{nn'}(t+\mathcal{T})$, reflects the influence of the laser field
with frequency $\Omega=2\pi/\mathcal{T}$.
The $L$ ideal leads are described by grand-canonical ensembles of electrons
at temperature $T$ and electro-chemical potential $\mu_\l$, $\l=1,\dots,L$.
Thus, the lead Hamiltonian reads
$H_\mathrm{leads} = \sum_{q\l} \epsilon_{q\l} \, c_{q\l}^\dagger c_{q\l}$,
where $c_{q\l}$ destroys an electron in state $q$ in lead $\l$.
All expectation values of lead operators can be traced back to
$\langle c_{q\l}^\dagger c_{q'\l'} \rangle = \delta_{qq'}
\delta_{\l\l'} f(\epsilon_{q\l}-\mu_\l)$, where $f(\epsilon)
=(1+\e^{\epsilon/k_B T})^{-1}$ denotes the Fermi function.
The model is completed by the molecule-leads tunnelling Hamiltonian
\begin{equation}
H_{\text{molecule-leads}}
=\sum_{q\l} V_{q\l} \, c_{q\l}^\dagger \, c_{\l} + \mathrm{h.c.}\ ,
\end{equation}
that connects each lead directly to one of the suitably labelled molecular
orbitals.  Since we are not interested here in the effects that arise from the
microscopic details of the molecule-lead coupling, we restrict our analysis in
the following to energy-independent couplings, \textit{i.e.} $\Gamma_{\l} = (2\pi/\hbar)
\sum_q |V_{q\l}|^2 \, \delta(\epsilon - \epsilon_{q\l})$.

\section{Perturbative description and Floquet ansatz}

Let us assume that the dynamics of the driven wire is dominated by the
time-dependent wire Hamiltonian so that the coupling to the leads can be taken
into account as a perturbation.  This allows to derive by use of standard
methods the approximate equation of motion for the total density operator
$\varrho(t)$, 
\begin{align}
\label{eq:master}
\dot\varrho(t) =
& -\frac{i}{\hbar}[H_{\rm molecule}(t)+H_{\rm leads},\varrho(t)] \\
& -\frac{1}{\hbar^2}\int\limits_0^\infty \!\mathrm{d}\tau
  [H_\mathrm{molecule-leads},[\widetilde{H}_\mathrm{molecule-leads}(t-\tau,t),
  \varrho(t)]] . \nonumber
\end{align}
We have omitted a transient term that depends purely on the initial preparation.
The tilde denotes operators in the interaction picture with respect to the
molecule plus the lead Hamiltonian, $\widetilde{O}(t,t')=U_0^\dagger(t,t') O(t)
U_0(t,t')$, where $U_0(t,t')$ is the time-evolution operator without the
coupling.  The first term describes the coherent dynamics of the electrons on
the wire, while the second term represents incoherent hopping of electrons
between the leads and the wire.

The net (incoming minus outgoing) electrical current that flows from lead $\l$
into the molecule is then given by the rate at which the electron number in the
corresponding lead decreases multiplied by the electron charge $-e$,
\begin{equation}
  \label{eq:current}
  I_\l(t) = e\, \frac{\mathrm{d}}{\mathrm{d}t}\langle N_\l\rangle\ .
\end{equation}
Note that this expectation value is time-dependent through the non-equilibrium
density operator $\varrho(t)$.  To evaluate the right-hand side of
Eq.~(\ref{eq:current}), we employ Eq.~\eqref{eq:master} to derive after some
algebra the result
\begin{equation}
\label{eq:current_general}
\begin{split}
I_\l(t)  = 
 -e   \Gamma_{\l} \bigg\{ &\mathop{\rm Re} 
\int\limits_0^\infty \frac{\mathrm{d}\tau}{\hbar} \! 
\int  \!\frac{\mathrm{d}\epsilon}{\pi}\,
e^{i\epsilon\tau/\hbar}          
f(\epsilon-\mu_\l)\big\langle[c_\l,
\tilde c_{\l}^\dagger(t-\tau,t)]_+\big\rangle \\
 &- \big\langle c_\l^\dagger
 c_{\l}\big\rangle\bigg\}\ .
\end{split}
\end{equation}
This expression still contains the yet unknown expectation values of solely
those wire operators with a direct connection to lead $\l$.  It depends in
particular on the Heisenberg operators $\tilde c_\l^\dagger(t-\tau,t)$ and thus
implicitly on the dynamics of the driven wire.  Let us therefore focus on the
single-particle dynamics of the periodically time-dependent wire Hamiltonian
$H_{nn'}(t)$.  An established procedure for the solution of the corresponding
Schr\"odinger equation is to employ a Floquet ansatz amounting to a
non-perturbative treatment of the external driving.  There one uses the fact
that a complete set of solutions is of the form
$|\Psi_\alpha(t)\rangle=\exp(-i\epsilon_\alpha t/\hbar)|\Phi_\alpha(t) \rangle$.
The so-called quasienergies $\epsilon_\alpha$ take over the role of the
energy eigenvalues in static systems and govern the long-time dynamics.  The Floquet
modes $|\Phi_\alpha(t)\rangle$ obey the time-periodicity of the driving field
which allows to express them as a Fourier series, $|\Phi_\alpha(t)\rangle =
\sum_{k=-\infty}^{\infty} \exp(-ik\Omega t) |\Phi_{\alpha,k}\rangle$.  They can
be obtained from the eigenvalue equation
\cite{Shirley1965a,Sambe1973a,Grifoni1998a}
\begin{equation}
\label{floquet_hamiltonian}
\Big(\sum_{n,n'}|n\rangle H_{nn'}(t) \langle n'|
-i\hbar\frac{\mathrm{d}}{\mathrm{d}t}\Big)
|\Phi_{\alpha}(t)\rangle = \epsilon_\alpha |\Phi_{\alpha}(t)\rangle
\ .
\end{equation}
Moreover, the Floquet modes define the complete set of operators
\begin{equation}
\label{eq:c_alpha}  
c_\alpha(t) = \sum_n \langle\Phi_\alpha(t)|n\rangle\, c_n  ,
\end{equation}
whose time-evolution assumes the convenient form $\tilde{c}_\alpha(t-\tau,t) =
\exp(i\epsilon_\alpha\tau/\hbar) c_\alpha(t)$.  The orthogonality of the Floquet
states at equal times \cite{Shirley1965a,Sambe1973a,Grifoni1998a} yields the
back-transformation $c_n=\sum_\alpha\langle n|\Phi_\alpha(t)\rangle$ and thus
results in the required spectral decomposition.
Using (\ref{eq:c_alpha}) and performing the energy and the $\tau$-integration
in Eq.~(\ref{eq:current_general}), we obtain for the time-averaged current
the main result
\begin{equation}
\label{eq:meancurrent}
\begin{split}
\bar{I}_\l = 
-\frac{e\Gamma_\l}{\hbar}\sum_{\alpha k}
\Big[&
\langle \Phi_{\alpha, k}|\l\rangle \langle \l|\Phi_{\alpha, k}\rangle 
  f(\epsilon_\alpha+k\hbar\Omega-\mu_\l) 
\\ &
-\sum_{\beta k'}
 \langle \Phi_{\alpha, k'+k} |\l\rangle\langle \l|\Phi_{\beta k'} \rangle
 R_{\alpha\beta,k}
\Big] .
\end{split}
\end{equation}
Here, we have introduced the expectation values $R_{\alpha\beta}(t)= \langle
c^\dagger_\alpha(t)c_{\beta}(t)\rangle$, which assume in the long-time limit the
time-periodicity of the driven system and thus can be decomposed into the
Fourier series $R_{\alpha\beta}(t)= \sum_k \exp(-ik\Omega t) R_{\alpha\beta,k}$.
It is straightforward to derive for the $R_{\alpha\beta,k}$ from the
density operator equation~(\ref{eq:master}) the following set of inhomogeneous linear
equations
\begin{align}
\label{eq:mastereq_fourier}
\lefteqn{\frac{i}{\hbar}(\epsilon_\alpha-\epsilon_\beta+k \hbar \Omega)R_{\alpha\beta,k}}
\\ && =
 \frac{1}{2}\sum_{\l k'} \Gamma_\l
    \Big\{ &
    \sum_{\beta'k''}
                \langle\Phi_{\beta,k''+k'}|\l\rangle
                \langle \l|\Phi_{\beta',k''+k}\rangle
                R_{\alpha\beta',k'}
\nonumber \\&& {}+{} &
    \sum_{\alpha'k''}
                \langle\Phi_{\alpha',k''+k'}|\l\rangle
                \langle \l|\Phi_{\alpha,k''+k}\rangle
                R_{\alpha'\beta,k'}
\nonumber \\ && {}-{} &
    \langle\Phi_{\beta,k'-k}|\l\rangle
    \langle \l|\Phi_{\alpha,k'}\rangle
    f(\epsilon_\alpha+k'\Omega-\mu_\l)
\nonumber \\ && {}-{} &
    \langle\Phi_{\beta,k'}|\l\rangle
    \langle \l|\Phi_{\alpha,k'+k}\rangle
    f(\epsilon_\beta+k'\Omega-\mu_\l)
   \Big\}
\nonumber\ ,
\end{align}
which will be solved numerically.  We have found that even in the case of
strong driving where the Floquet states comprise many sidebands, a few Fourier
coefficients $R_{\alpha\beta,k}$ are in fact sufficient to obtain numerical
convergence.  This justifies \textit{a posteriori} the use
of the Floquet states as a basis set.

To conclude the technical part of this work, we note that our approach goes
beyond a linear response treatment of the driving and additionally does not use
a so-called rotating wave approximation (RWA) \cite{Blumel1991a,Bruder1994a},
which neglects the oscillatory contributions to $R_{\alpha\beta}(t)$ by the
ansatz $R_{\alpha\beta,k}=P_\alpha\,\delta_{\alpha,\beta}\,\delta_{k,0}$.  In
fact, we found that a RWA solution delivers inaccurate results in the vicinity
of quasienergy degeneracies.

%%%%%%%%%%%%%%%%%%%%%%%%%%%%%%%%%%%%%%%%%%%%%%%%%%%%%%%%%%%%%%%%%%%%%%%%%%%%%%%%
\section{Optical current gate}

As a first setup that may be suitable as a current control device, we
investigate the transport through a two-level system, \textit{i.e.} a wire that consists
of $N=2$ sites --- one of them is coupled to the left lead and the other to
the right lead.  Then the time-dependent wire Hamiltonian reads in the basis of
the molecular orbitals
\begin{equation}
  \label{eq:twosites}
    H_\mathrm{molecule}(t) = 
    \begin{pmatrix}
      E_\mathrm{L} && -\Delta\\
      -\Delta      && E_\mathrm{R}
    \end{pmatrix}
    + A 
    \begin{pmatrix}
      1 && 0\\
      0 && -1
    \end{pmatrix}\cos(\Omega t)\ ,
\end{equation}
where $\Delta$ denotes the tunnel matrix element between the two sites and
$E_\mathrm{L}$ and $E_\mathrm{R}$ are the corresponding on-site energies.  The
laser field contributes to the Hamiltonian (\ref{eq:twosites}) a time-dependent
bias with amplitude $A=-e\mathcal{E}d$, \textit{i.e.} charge times electrical
field strength times the site-to-site distance.  Note that the electrical field
may be drastically enhanced due to the presence of the metallic tips
\cite{Demming1998a}.  The effective coupling to each lead is assumed to be
equal, $\Gamma_{\l} = \Gamma$, and an external voltage $V$ is taken into account
by a difference in the electro-chemical potentials,
$\mu_\mathrm{L}-\mu_\mathrm{R}=-eV$.

We use in all numerical calculations the tunnel matrix element $\Delta$ as the
energy unit and assume that the effective couplings to the leads are by one order of
magnitude smaller, $\hbar\Gamma=0.1\Delta$. This corresponds to
a large contact resistance and ensures the applicability of a perturbational
approach.  A realistic value is $\Delta=0.1\mathrm{eV}$, resulting in a current
unit $e\Gamma=0.256\,\mathrm{\mu A}$.  For a site-to-site distance of $10${\AA}
and a laser frequency $\Omega=10\Delta/\hbar$, the driving amplitudes
considered below correspond to an electrical field amplitude of
$10^6\,\mathrm{V}/\mathrm{cm}$ at $1\,\mu\mathrm{m}$ wavelength.

%%%% Figure 2 %%%%%%%%%%%%%%%%%%%%%%%%%%%%%%%%%%%%%%%%%%%%%%%%%%%%%%%%%%%%%%%
\begin{figure}
  \begin{center}
    \includegraphics[width=7.5truecm]{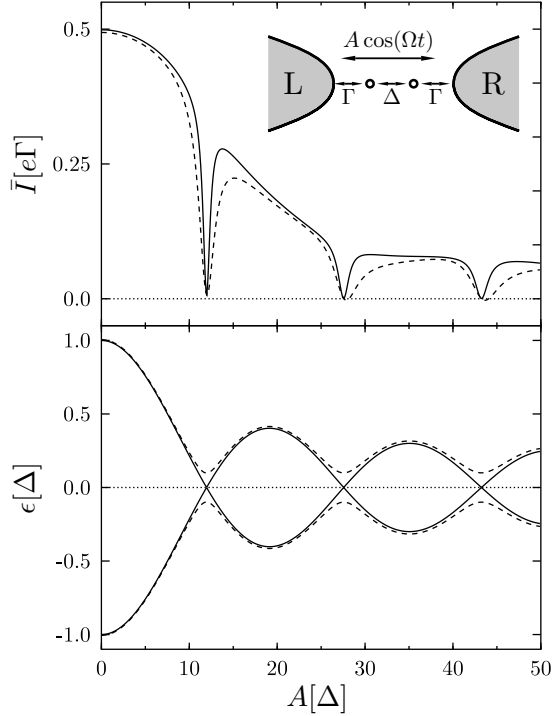}
  \end{center}
  \caption{Average current and quasienergy spectrum versus driving amplitude
    for a wire which consists of two sites between two electrodes (cf.\ inset)
    for unbiased ($E_\mathrm{R}=E_\mathrm{L}=0$, solid lines) and
    biased ($E_\mathrm{R}=-E_\mathrm{L}=0.1\Delta$, dashed lines)
    on-site energies.
    The leads' chemical potentials are $\mu_\mathrm{R}=-\mu_\mathrm{L}=10 \Delta$;
    the other parameters read $\hbar\Omega=10\Delta$, $k_B T=0.25\Delta$, 
    $\hbar\Gamma=0.1\Delta$.
    }
  \label{fig:twolevel}
\end{figure}%
%%%% Figure 2 %%%%%%%%%%%%%%%%%%%%%%%%%%%%%%%%%%%%%%%%%%%%%%%%%%%%%%%%%%%%%%%
The time-averaged current $\bar{I}=\bar{I}_\mathrm{L}=-\bar{I}_\mathrm{R}$
through the molecule in a case where both on-site energies are equal is
depicted in Fig.~\ref{fig:twolevel}.  As a striking feature, we find that at
certain values of the driving amplitude, the current collapses to less than 1\%
of its maximal value reached in the absence of the driving.  Closer inspection
(not shown) reveals that the width of this depression is proportional to the
molecule-lead coupling $\Gamma$.  Comparison with the quasienergy spectrum in
the lower panel demonstrates that the current break downs occur at quasienergy
crossings.  This relates the present phenomenon to the CDT, \textit{i.e.} the
standstill of the tunnel dynamics in a driven bistable potential at quasienergy
crossings \cite{Grossmann1991a}.  For the \textit{isolated} two-level system
(\ref{eq:twosites}) with $\Delta\ll\Omega,A$, the CDT condition reads
$\mathrm{J}_0(2A/\hbar\Omega)=0$ \cite{Grossmann1992a}, \textit{i.e.} the suppression
of the tunnelling dynamics is related to the zeros of the Bessel function
$\mathrm{J}_0$.  As our analysis reveals, the same condition results in a
suppression of the transport through the \textit{open} system.

An external voltage may be of peculiar influence to the on-site energies of a
molecular wire \cite{Mujica2000a} and cause an effective bias $E_\mathrm{L}\neq
E_\mathrm{R}$ in originally symmetric molecules.  Thus, a crucial question is
whether the above current suppression is stable against such a modification.
The broken lines in Fig.~\ref{fig:twolevel} demonstrate that this is indeed the
case.  Although the quasienergies now form an avoided crossing, the current
breakdowns do survive; they are even more pronounced, but slightly shifted
towards larger driving amplitudes.  This robustness of CDT based current
control combined with the huge on/off ratio suggests the presented setup as a
promising alternative to structural chemistry-based switching devices
\cite{Chen1999a,Collier2000a}.

%%%%%%%%%%%%%%%%%%%%%%%%%%%%%%%%%%%%%%%%%%%%%%%%%%%%%%%%%%%%%%%%%%%%%%%%%%%%%%%%
\section{Optical current router}

\begin{figure}
  \begin{center}
    \includegraphics[width=7.5truecm]{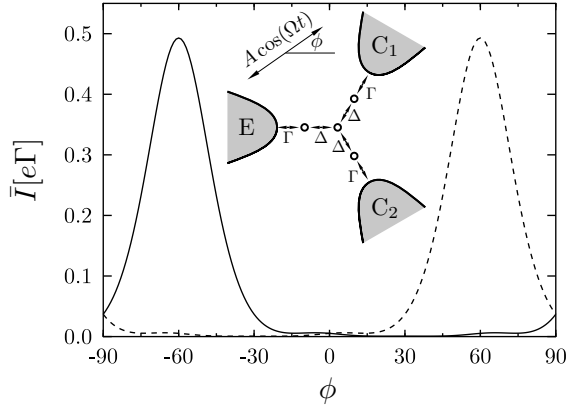}
  \end{center}
  \caption{Average currents through contacts $\mathrm{C}_1$ (solid)
    and $\mathrm{C}_2$ (broken) as a function of the polarisation angle $\phi$
    for the three-terminal device depicted in the inset.  The chemical
    potentials are $\mu_\mathrm{E}=-\mu_{\mathrm{C}_1} =-\mu_{\mathrm{C}_2}=50
    \Delta$; the on-site energies $E_n=0$.  The driving field is specified by
    the strength $A=25\Delta$ and the angular frequency $\Omega=10\Delta/\hbar$;
    the effective coupling is $\hbar\Gamma=0.1\Delta$ and the temperature $k_B
    T=0.25\Delta$.  The maximal value of the current ratio
    $I_{\mathrm{C}_1}/I_{\mathrm{C}_2}\approx 100$ is assumed at $\phi=60^\circ$. }
  \label{fig:switch1}
\end{figure}

An experimentally more ambitious configuration consists in a planar molecule
with $N=4$ sites, three of which are coupled to a central site and are
directly connected to leads (cf.\ inset of Fig.~\ref{fig:switch1}).  We borrow
from electrical engineering the designation $\mathrm{E}$, $\mathrm{C}_1$, and
$\mathrm{C}_2$.  Here, an external voltage is always applied such that
$\mathrm{C}_1$ and $\mathrm{C}_2$ have equal electro-chemical potential,
\textit{i.e.} $\mu_{\mathrm{C}_1}=\mu_{\mathrm{C}_2}\neq\mu_\mathrm{E}$.  In a perfectly
symmetric molecule, where all on-site energies equal each other, reflection
symmetry at the horizontal axis ensures that any current which enters at
$\mathrm{E}$, is equally distributed among $\mathrm{C}_{1,2}$, thus
$I_{\mathrm{C}_1}=I_{\mathrm{C}_2}=-I_\mathrm{E}/2$.
Since this structure in Fig.~\ref{fig:switch1} is essentially two-dimensional,
we have to take also the polarisation of the laser field into account.  We
assume it to be linear with an polarisation angle $\phi$ as sketched in the
inset of Fig.~\ref{fig:switch1}.  The effective driving amplitudes of the
orbitals which are attached to the leads acquire now a geometric factor which
is only the same for both orbitals $\mathrm{C}_1$ and $\mathrm{C}_2$ when
$\phi=0$.  For any other polarisation angle, the mentioned symmetry is broken
and the outgoing currents may be different from each other.  The difference may
be huge, as depicted in Fig.~\ref{fig:switch1}.  Their ratio varies from unity
for $\phi=0$ up to the order of 100 for $\phi=60^\circ$.  Thus, adapting the
polarisation angle enables one to route the current towards the one or the other
drain.
%%%% Figure 3 %%%%%%%%%%%%%%%%%%%%%%%%%%%%%%%%%%%%%%%%%%%%%%%%%%%%%%%%%%%%%%%
%%%% Figure 3 %%%%%%%%%%%%%%%%%%%%%%%%%%%%%%%%%%%%%%%%%%%%%%%%%%%%%%%%%%%%%%%

\begin{figure}
  \begin{center}
    \includegraphics[width=7.5truecm]{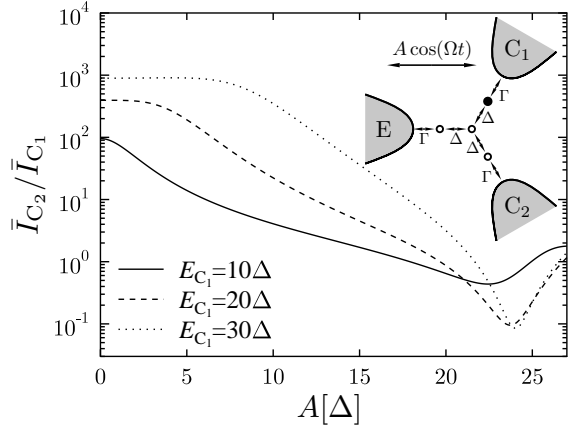}
  \end{center}
  \caption{Ratio of the outgoing average currents versus driving strength $A$
    for the three-terminal device at a polarisation angle $\phi=0$.
    The filled circle in the inset depicts a site with an on-site energy
    $E_{\mathrm{C}_1}$ that differs from the others.
    All other on-site energies and parameters as in Fig.~\ref{fig:switch1}.
    }
  \label{fig:switch2}
\end{figure}

Alternatively, one can keep the polarisation angle at $\phi=0$ and break the 
reflection symmetry by using an intrinsically asymmetric molecule, as
sketched in the inset of Fig.~\ref{fig:switch2}. This allows to 
control sensitively the ratio of the outgoing currents by the strength $A$ of the
external field, cf.\ Fig.~\ref{fig:switch2}.  The switching range comprises
up to four orders of magnitude with an exponential sensitivity.
%%%% Figure 4 %%%%%%%%%%%%%%%%%%%%%%%%%%%%%%%%%%%%%%%%%%%%%%%%%%%%%%%%%%%%%%%
%%%% Figure 4 %%%%%%%%%%%%%%%%%%%%%%%%%%%%%%%%%%%%%%%%%%%%%%%%%%%%%%%%%%%%%%%

%%%%%%%%%%%%%%%%%%%%%%%%%%%%%%%%%%%%%%%%%%%%%%%%%%%%%%%%%%%%%%%%%%%%%%%%%%%%%
\section{Concluding remarks}

We have presented a method for the efficient numerical computation of currents
through periodically time-dependent networks with two or more contacts to
external leads.  The application to two types of setups substantiated that
external fields bear a wealth of possibilities for the manipulation of
electrical currents: in a molecular wire the current can be suppressed by
proper time-dependent fields.  In a three terminal device, it is possible to
route by tailored optical fields the current that enters from a source towards
the one or the other drain.

The authors hope that their proposals will motivate experimentalists to accept
the challenge of implementing the proposed molecular wire schemes in the
laboratory.  The two-terminal current gate can possibly be realized using break
junctions exposed to a laser field.  Alternatively, one could use a
self-assembled, laser-irradiated maze-like layer of sparsely distributed
conducting molecules on a conducting surface. Then by positioning a scanning
tunnelling microscope tip directly over one such molecule, it should be
possible to measure the features of the predicted gating behaviour.
Experimentally more ambitious is the realization of the arrangement in
Fig.~\ref{fig:switch1} with a planar molecule contacted to three electrodes.
Here again, laser-irradiated self-assemblies of molecules such as carbon
nanotube complexes or of biomolecules like metalised DNA \cite{Braun1998a}, or
the use of cationic lipid-DNA complexes \cite{Raedler1998a} as DNA-nanocables,
with the centre-molecule covalently bound to such planar structures, might make
the experiment feasible.

A completely different realization of our findings should be possible in
semiconductor heterostructures. There, instead of a molecule, coherently
coupled quantum dots \cite{Blick1996a} form the central system. Furthermore,
owing to the lower level spacings, the suitable frequency of the coherent
radiation source is then in the microwave spectral range.

%%%%%%%%%%%%%%%%%%%%%%%%%%%%%%%%%%%%%%%%%%%%%%%%%%%%%%%%%%%%%%%%%%%%%%%%%%%%%
\ack
We appreciate helpful discussions with Igor Goychuk and Gert-Ludwig Ingold.
This work has been supported by the Deutsche Forschungsgemeinschaft
through SFB 486  and by the Volkswagenstiftung under grant No.\ I/77 217.


\begin{thebibliography}{10}
\expandafter\ifx\csname url\endcsname\relax
  \def\url#1{\texttt{#1}}\fi
\expandafter\ifx\csname urlprefix\endcsname\relax\def\urlprefix{URL }\fi

\bibitem{Reed1997a}
M.~A. Reed, C.~Zhou, C.~J. Muller, T.~P. Burgin, J.~M. Tour, Conductance of a
  molecular junction, Science 278 (1997) 252.

\bibitem{Cui2001a}
X.~D. Cui, A.~Primak, X.~Zarate, J.~Tomfohr, O.~F. Sankey, A.~L. Moore, T.~A.
  Moore, D.~Gust, G.~Harris, S.~M. Lindsay, Reproducible measurement of
  single-molecule conductivity, Science 294 (2001) 571.

\bibitem{Joachim2000a}
C.~Joachim, J.~K. Gimzewski, A.~Aviram, Electronics using hybrid-molecular and
  mono-layer devices, Nature 408 (2000) 541.

\bibitem{Schon2001a}
J.~H. Sch\"on, H.~Meng, Z.~Bao, Self-assembled monolayer organic field-effect
  transistors, Nature 413 (2001) 713.

\bibitem{Reichert2002a}
J.~Reichert, R.~Ochs, D.~Beckmann, H.~Weber, M.~Mayor, H.~v.~Loehneysen,
  Driving current through single organic molecules, Phys. Rev. Lett. 88 (2002)
  176804.

\bibitem{Nitzan2001a}
A.~Nitzan, Electron transmission through molecules and molecular interfaces,
  Ann. Rev. Phys. Chem. 52 (2001) 681.

\bibitem{Mujica1994a}
V.~Mujica, M.~Kemp, M.~A. Ratner, Electron conduction in molecular wires. i. a
  scattering formalism, J. Chem. Phys. 101 (1994) 6849.

\bibitem{Nitzan2001b}
A.~Nitzan, A relationship between electron-transfer rates and molecular
  conduction, J. Phys. Chem. A 105 (2001) 2677.

\bibitem{Petrov2001a}
E.~G. Petrov, P.~H\"anggi, Nonlinear electron current through a short molecular
  wire, Phys. Rev. Lett. 86 (2001) 2862.

\bibitem{Heurich2001a}
J.~Heurich, J.~C. Cuevas, W.~Wenzel, G.~Sch\"on, Electrical transport through
  single-molecule junctions: from molecular orbitals to conduction channels,
  cond-mat/  (2001) 0110147.

\bibitem{Grossmann1991a}
F.~Grossmann, T.~Dittrich, P.~Jung, P.~H\"anggi, Coherent destruction of
  tunneling, Phys. Rev. Lett. 67 (1991) 516.

\bibitem{Grossmann1992a}
F.~Gro{\ss}mann, P.~H\"anggi, Localization in a driven two-level dynamics,
  Europhys. Lett. 18 (1992) 571.

\bibitem{Morillo1993a}
M.~Morillo, R.~I. Cukier, Control of proton transfer reactions with external
  fields, J. Chem. Phys. 98 (1993) 4548.

\bibitem{Goychuk1996a}
I.~A. Goychuk, E.~G. Petrov, V.~May, Control of the dynamics of a dissipative
  two-level system by a strong periodic field, Chem. Phys. Lett. 253 (1996)
  428.

\bibitem{Grifoni1998a}
M.~Grifoni, P.~H\"anggi, Driven quantum tunneling, Phys. Rep. 304 (1998) 229.

\bibitem{Shirley1965a}
J.~H. Shirley, Solution of the {S}chr\"odinger equation with a {H}amiltonian
  periodic in time, Phys. Rev. 138 (1965) B979.

\bibitem{Sambe1973a}
H.~Sambe, Steady states and quasienergies of a quantum-mechanical system in an
  oscillating field, Phys. Rev. A 7 (1973) 2203.

\bibitem{Lehmann2002b}
J.~Lehmann, S.~Kohler, P.~H\"anggi, A.~Nitzan, Molecular wires acting as
  quantum ratchets, Phys. Rev. Lett. 88 (2002) 228305.

\bibitem{Blumel1991a}
R.~Bl\"umel, A.~Buchleitner, R.~Graham, L.~Sirko, U.~Smilansky, H.~Walter,
  Dynamical localisation in the microwave interaction of {R}ydberg atoms: The
  influence of noise, Phys. Rev. A 44 (1991) 4521.

\bibitem{Bruder1994a}
C.~Bruder, H.~Schoeller, Charging effects in ultrasmall quantum dots in the
  presence of time-varying fields, Phys. Rev. Lett. 72 (1994) 1076.

\bibitem{Demming1998a}
F.~Demming, J.~Jersch, K.~Dickmann, P.~I. Geshev, Calculation of the field
  enhancement on laser-illuminated scanning probe tips by the boundary element
  method, Appl. Phys. B 66 (1998) 593.

\bibitem{Mujica2000a}
V.~Mujica, A.~E. Roitberg, M.~Ratner, Molecular wire conductance: Electrostatic
  potential spatial profile, J. Chem. Phys. 112 (2000) 6834.

\bibitem{Chen1999a}
J.~Chen, M.~A. Reed, A.~M. Rawlett, J.~M. Tour, Large on-off ratios and
  negative differential resistance in a molecular electronic device, Science
  286 (1999) 1550.

\bibitem{Collier2000a}
C.~P. Collier, G.~Mattersteig, E.~W. Wong, Y.~Luo, K.~Beverly, J.~Sampaio,
  F.~M. Raymo, J.~F. Stoddart, J.~R. Heath, A [2]catenane-based solid state
  electronically reconfigurable switch, Science 289 (2000) 1172.

\bibitem{Braun1998a}
E.~Braun, Y.~Eichen, U.~Sivan, G.~Ben-Yoseph, {DNA}-templated assembly and
  electrode attachment of a conducting silver wire, Nature 391 (1998) 775.

\bibitem{Raedler1998a}
O.~J. R\"adler, I.~Koltover, A.~Jamieson, T.~Salditt, C.~R. Safinya, Structure
  and interfacial aspects of self-assembled cationic lipid-{DNA} gene carrier
  complexes, Langmuir 14 (1998) 4272.

\bibitem{Blick1996a}
R.~H. Blick, R.~J. Haug, J.~Weis, D.~Pfannkuche, K.~v. Klitzing, K.~Eberl,
  Single-electron tunneling through a double quantum dot: The artificial
  molecule, Phys. Rev. B 53 (1996) 7899.

\end{thebibliography}
\end{document}